\documentclass{aa}
\usepackage{epsfig}

\def\etal{{\rm et al.\thinspace}}
\def\eg{{\rm e.g.\ }}

\def\ie{{\rm i.e.\ }}
\def\cf{{\rm cf.\ }}

\def\Mdot{\hbox{${\dot M}$} \,}

\def\h50{\hbox{$h_{50}$\,}}
\def\spose#1{\hbox to 0pt{#1\hss}}
\def\ltsimm{\mathrel{\spose{\lower 3pt\hbox{$\sim$}}
	\raise 2.0pt\hbox{$<$}}}
\def\ltsim{$\mathrel{\spose{\lower 3pt\hbox{$\sim$}}
	\raise 2.0pt\hbox{$<$}}$}
\def\gtsimm{\mathrel{\spose{\lower 3pt\hbox{$\sim$}}
	\raise 2.0pt\hbox{$>$}}}
\def\gtsim{$\mathrel{\spose{\lower 3pt\hbox{$\sim$}}
	\raise 2.0pt\hbox{$>$}}$}

\def\fract#1/#2{\leavevmode\kern.1em                   % e.g. \fract 10/3
   \raise.5ex\hbox{\the\scriptfont0 #1}\kern-.1em
   /\kern-.15em\lower.25ex\hbox{\the\scriptfont0 #2}}

\def\cm{{\rm\thinspace cm}}
\def\erg{{\rm\thinspace erg}}

\def\km{{\rm\thinspace km}}

\def\Lsol{\hbox{${\rm\thinspace L_{\odot}}$}}

\def\Msol{\hbox{${\rm\thinspace M_{\odot}}$}}

\def\pc{{\rm\thinspace pc}}
\def\s{{\rm\thinspace s}}

\def\erg{{\rm\thinspace erg}}

\def\cm2{\hbox{${\rm\cm^{2}\,}$}}
\def\pcm2{\hbox{${\rm\cm^{-2}\,}$}}
\def\ergpcm3ps{\hbox{${\rm\erg\cm^{-3}\s^{-1}\,}$}}

\def\kmps{\hbox{${\rm\km\s^{-1}\,}$}}

\def\Lsolppc3{\hbox{${\rm\Lsol\pc^{-3}\,}$}}
\def\Msolppc3{\hbox{${\rm\Msol\pc^{-3}\,}$}}

\begin{document}
   
\title{Self-Similar Evolution of Wind-Blown Bubbles with Mass-loading 
by Conductive Evaporation}

\author{J.M. Pittard, J.E. Dyson \and T.W. Hartquist}

\offprints{J.M. Pittard (jmp@ast.leeds.ac.uk)}

\institute{Department of Physics \& Astronomy, The University of Leeds, 
        Woodhouse Lane, Leeds, LS2 9JT, United Kingdom\\}

\date{Received ...; accepted...}

\thesaurus{08     % A&A Section 6: Diffuse matter in space
              (02.08.1;  % Hydrodynamics
               02.19.1;  % Shock waves,
               08.13.2;  % Stars: mass-loss,
               09.02.1;  % ISM: bubbles,
               11.01.2)} % Galaxies: active

\titlerunning{Self-similar evolution of Mass-Loaded WBBs}
\authorrunning{Pittard, Dyson \& Hartquist}

%\begin{document}

\maketitle
\label{firstpage}

\begin{abstract}
We present similarity solutions for adiabatic bubbles that are blown by
winds having time dependent mechanical luminosities and that are each
mass-loaded at a rate per unit volume proportional to $T^{5/2} r^{\lambda}$,
where $T$ is the temperature, $r$ is the distance from the bubble center,
and $\lambda$ is a constant. In the limit of little mass loading a 
similarity solution found by Dyson (\cite{D1973}) for expansion into a
smooth ambient medium is recovered. For solutions that give the mass of
swept-up ambient gas to be less than the sums of the masses of the wind
and the evaporated material, $\lambda \gtsimm 4$. The Mach number in
a shocked mass-loaded wind shows a radial dependence that varies qualitatively
from solution to solution. In some cases it is everywhere less than unity in
the frame of the clumps being evaporated, while in others it is everywhere
greater than unity. In some solutions the mass-loaded shocked wind undergoes
one or two sonic transitions in the clump frame. Maximum possible values
of the ratio of evaporated mass to stellar wind mass are found as a 
consequence of the evaporation rates dependence on temperature and the
lowering of the temperature by mass-loading. Mass-loading tends 
to reduce the emissivity in the interior of the
bubble relative to its limb, whilst simultaneously increasing the central
temperature relative to the limb temperature.
\end{abstract}

\section{Introduction}
\label{sec:intro}
Many authors have pointed out that mass pick-up from embedded clumps
can significantly modify the structures of a wide variety of astrophysical
flows. These include, for example, McKee \& Ostriker (\cite{MO1977}) in
their theory of the interstellar medium, Hartquist \etal (\cite{HDPS1986}) in
the study of wind blown nebulae, and Chi\`{e}ze \& 
Lazareff (\cite{CL1981}), Cowie, McKee \& Ostriker (\cite{CMO1981}) 
and Dyson \& Hartquist (\cite{DH1987}) in the study of supernova remnants 
(SNRs).

Observations of wind-blown
bubbles (WBBs) (\eg Smith \etal \cite{SPDH1984}; Meaburn \etal 
\cite{MNBDW1991}) have led to the conclusion that the interactions of the
winds with clumps of stellar material ejected during prior stages of mass-loss
greatly affect the structure and evolution of the bubbles (see Williams,
Hartquist \& Dyson \cite{WHD1995} and references therein). Consequently, 
a number of numerical investigations of this process have been conducted.
Detailed one-dimensional, time dependent hydrodynamic models of specific 
WBBs have been constructed by Arthur, Dyson \& Hartquist (\cite{ADH1993},
\cite{ADH1994}) and Arthur, Henney \& Dyson (\cite{AHD1996}). 
Numerical solutions for the Mach numbers of steady
mass-loaded winds have been presented by Williams \etal (\cite{WHD1995}),
and studies of steady hydromagnetic mass-loaded winds have been performed
by Williams, Dyson, \& Hartquist (\cite{WDH1999}).

In this paper we derive similarity solutions for the structures and 
evolution of mass-loaded WBBs. This work complements that on similarity 
solutions for SNRs obtained by Chi\`{e}ze \& Lazareff (\cite{CL1981}) 
and Dyson \& Hartquist (\cite{DH1987}). Here we consider a constant 
{\em rate} of energy input, $\dot{E}$, rather than a constant {\em total} 
energy, and assume that the mass-loading occurs due to conductively
driven evaporation. The solutions are potentially relevant to WBBs created
by a fast wind interacting with a clumpy AGB superwind, by the wind of a 
young high-mass star interacting with surrounding molecular material, and
the wind of an active galactic nucleus impacting its environment.

\section{Similarity Solutions}
\label{sec:sim_solutions}

\subsection{Overview of assumptions}
\label{sec:assumptions}
The evaporation of spherical clouds by hot gas has been analyzed by
Cowie \& McKee (\cite{CM1977}) and McKee \& Cowie (\cite{MC1977}).
If the mean free path for electron-electron energy change in
the ambient gas is less than about half of the radius of a clump, the
mass evaporation rate from an individual clump scales as $T^{5/2}$,
where $T$ is the temperature of the material surrounding the clump. 
For a typical jump of $1000\kmps$ at the reverse shock (corresponding to
a post-shock temperature of $T \approx 1.4 \times 10^{7}$~K), and a clump 
electron number density of $n_{e} \approx 10^{4}$, the mean free path for
electron energy exchange is $\lambda \approx 5 \times 10^{15}$~cm. This
is somewhat smaller than the lengthscale of a clump (\eg in a planetary
nebula, Meaburn \etal \cite{MCBWHS1998}), so the assumption of unsaturated 
conduction with a $T^{5/2}$ temperature dependence of $\dot{\rho}$, 
the mass evaporation rate per unit volume, should be reasonable throughout
most of the mass-loading region if mass-loading is significant.

We also assume that the clumps are 
stationary with respect to the star, and that they are also cold and dense.
Therefore the clumps have neither kinetic or internal energy, and the 
picked-up mass is thus injected at zero velocity. The stripped mass 
then acquires momentum and energy from the general more tenuous flow. Hence
the only source term in the hydrodynamic equations is $\dot{\rho}$. In
addition we assume that at any given point the injected mass reaches
the general flow velocity and temperature instantaneously \ie that the 
characteristic scale length of the mixing region is much smaller than the 
dimensions of the bubble. We also assume that the clumps are sufficiently 
numerous that they can be considered continuously distributed such that
a hydrodynamic treatment applies. We suppose that 
conductive energy transport can be neglected in the energy equation 
describing the global evolution of the WBB. The WBBs are assumed to evolve 
adiabatically, whilst the swept-up interclump medium is assumed to 
catastrophically cool into a thin dense shell. The clumps are assumed 
to pass through this shell without any interaction, and then on into 
the region of shocked stellar wind. The unshocked stellar 
wind is assumed to have a flow speed much greater than its thermal speed.
Hence its thermal energy is negligible with respect to its kinetic energy,
and since no mass-loading occurs in this region, we ignore the energy
equation for this part of the flow. 
Finally we assume that any external 
forces, such as gravity or radiation pressure, are negligible.

\begin{figure}  
\epsfig{figure=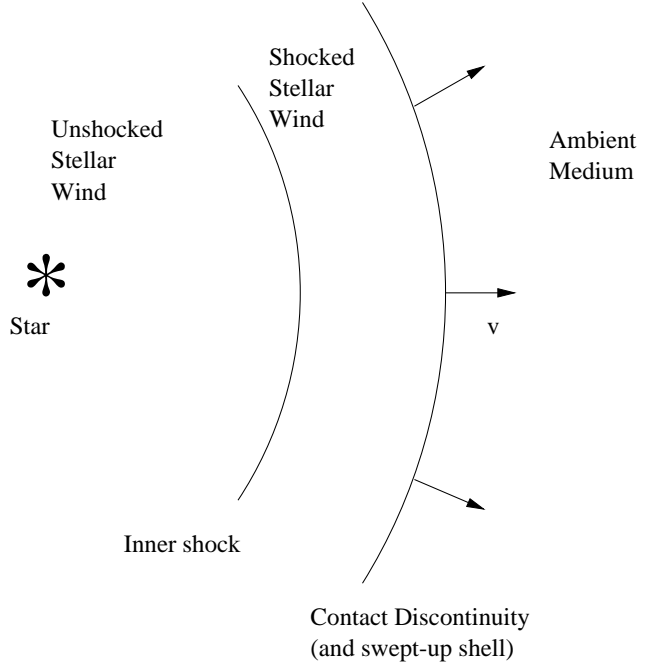,width=8.8cm,angle=-90}
\caption[]{Schematic of the flow structure
in our solutions. The central star blows out a bubble into the surrounding 
medium which is characterized by a region of unshocked stellar wind (USW), a
region of shocked stellar wind (SSW), and a dense shell of swept-up 
material (SUM). We assume that the shocked ambient medium quickly 
radiates all of its internal energy, whilst the cooling time of the SSW
is assumed to be much longer than the current age of the bubble. 
Mass-loading only occurs in the SSW.}
\label{fig:schematic}
\end{figure}
 
A schematic of the flow topology is shown in Fig.~\ref{fig:schematic}.
The hydrodynamical equations for the SSW region are:

\begin{equation}
\frac{\partial \rho}{\partial t} + \nabla \cdot (\rho u) = \dot{\rho}
\end{equation}

\begin{equation}
\frac{\partial (\rho u)}{\partial t} + \nabla \cdot (\rho u^{2}) 
 + \frac{\partial P}{\partial r} = 0 
\end{equation}

\begin{equation}
\frac{\partial (\frac{1}{2} \rho u^{2} + \varepsilon)}{\partial t} 
 + \nabla \cdot (\rho u(\frac{1}{2} u^{2} + \frac{5}{3} 
 \frac{\varepsilon}{\rho})) = 0 
\end{equation}

\noindent where the symbols have their usual meanings. For conductively 
induced mass-loading, the evaporation of the clumps only occurs in the 
SSW region. In the USW region we assume no interaction between the clumps 
and the fast stellar wind.

\subsection{The Similarity Variables}
\label{sec:evap}
Let the interclump medium have a density of the form 
$\rho_{{\rm ic}} = \rho_{0} \, r^{\beta}$, and let us consider the case 
where the mass-loading rate is also radially dependent:
$\dot{\rho} = Q \, (\varepsilon/\rho)^{5/2} \, r^{\lambda}$. 
The problem as its stands has five parameters with independent dimensions, 
namely $r$, $t$, $Q$, $\rho_{0}$, and the rate of energy input by the 
stellar wind
$\dot{E}$. The solution inside the bubble depends on $r$, $t$, $Q$, and
$\dot{E}$, whilst the solution ahead of the contact discontinuity 
depends on $r$, $t$, $\rho_{0}$, and $\dot{E}$. Inside the bubble we
can construct a variable $r = x \, {\dot{E}}^{u} \, t^{v} \, Q^{w}$ with
a dimension of length. Outside the bubble we construct a similar
variable $r' = x' \, \dot{E}^{x} \, t^{y} \, \rho_{0}^{z}$. Both $x$ and $x'$
are dimensionless. For a similarity solution to exist $v=y$, which demands 
that $\lambda = (5 + 7\beta)/3$. This reduces the number of dimensional
parameters to 4: $r$, $t$, $Q$, and $\dot{E}$.
The solution may then be expressed in terms of the dimensionless 
similarity variables $x$, $f(x)$, $g(x)$, and $h(x)$ where:

\begin{equation}
\label{eq:sim_r}
r = x \, Q^{-1/(10 + \lambda)} \, \dot{E}^{1/(10 + \lambda)} \, 
t^{7/(10 + \lambda)}
\end{equation}

\begin{equation}
\label{eq:sim_rho}
\rho = Q^{5/(10 + \lambda)} \, \dot{E}^{(5 + \lambda)/(10 + \lambda)}
       \, t^{-(5 - 3\lambda)/(10 + \lambda)} \, f(x)
\end{equation}

\begin{equation}
\label{eq:sim_u}
u    = Q^{-1/(10 + \lambda)} \, \dot{E}^{1/(10 + \lambda)} \,
       t^{-(3 + \lambda)/(10 + \lambda)} \, g(x)
\end{equation}

\begin{equation}
\label{eq:sim_eps}
\varepsilon = Q^{3/(10 + \lambda)} \, 
       \dot{E}^{(7 + \lambda)/(10 + \lambda)} \,
       t^{-(11 - \lambda)/(10 + \lambda)} \, h(x)
\end{equation}

\noindent The hydrodynamic equations for the region of shocked stellar 
wind become the following set of coupled ordinary differential equations:

\begin{eqnarray}
\label{eq:ode1_evap}
& \left(g - \frac{7x}{10 + \lambda}\right)f' + fg' & + \frac{2fg}{x} - 
\frac{5 - 3\lambda}{10 + \lambda}f \nonumber \\  
 & & - \left(\frac{h}{f}\right)^{5/2} x^{\lambda} = 0 \\
\label{eq:ode2_evap}
& \left(g - \frac{7x}{10 + \lambda}\right)g' + \frac{2}{3f}h' & - 
\frac{3 + \lambda}{10 + \lambda}g \nonumber \\ 
 & & + \frac{g}{f}\left(\frac{h}{f}\right)^{5/2} x^{\lambda} = 0 \\
& \left(g - \frac{7x}{10 + \lambda}\right)h' + \frac{5}{3}hg' & - 
\label{eq:ode3_evap}
\frac{11 - \lambda}{10 + \lambda}h \nonumber \\ 
 & & + \frac{10}{3x}hg - \frac{g^{2}}{2}\left(\frac{h}{f}\right)^{5/2} 
x^{\lambda} = 0 
\end{eqnarray}

\noindent where a prime denotes derivation with respect to x. It is 
simple to rearrange these equations to find $f'$, $g'$, and $h'$. 

\subsubsection{Boundary conditions}
\label{sec:bound_con}
We start our integration at a point very close to the star. As 
$r \rightarrow 0$, $4 \pi r^{2} \, \rho v \rightarrow \dot{M}_{{\rm c}}$, 
where $\dot{M}_{{\rm c}}$ is the {\em current} rate of mass-loss of the star 
($\dot{M}$ is a function of time if $\lambda \neq -3$). Substituting 
the similarity variables we obtain:

\begin{equation}
\label{eq:mdotc}
\dot{M}_{{\rm c}} = \lim_{x \rightarrow 0} 4 \pi \, x^{2} \, 
f g \, \Xi
\end{equation}

\noindent where $\Xi = Q^{2/(10+\lambda)} \,
\dot{E}^{(8+\lambda)/(10+\lambda)} \, t^{(6+2\lambda)/(10+\lambda)}$. 
We define $\phi$, such that

\begin{equation}
\label{eq:phi}
\phi = \frac{\Mdot_{{\rm c}}} \Xi = \lim_{x \rightarrow 0}
4 \pi \, x^{2} \, f g
\end{equation}

\noindent which is a measure of the ratio of evaporated mass to wind mass.
We also require that as $r \rightarrow 0$, $2 \pi r^{2} \,
\rho \, v^{3} \rightarrow \dot{E}$. \ie that

\begin{equation}
\label{eq:edot}
\dot{E} = \lim_{x \rightarrow 0} 2 \pi \, x^{2} \, 
f g^{3} \, \dot{E}
\end{equation}

\noindent Cancelling the $\dot{E}$'s and substituting into Eq.~\ref{eq:phi}
we obtain the values of $f$ and $g$ at $x=\Delta x$, the inner boundary of our
integration: $g = \sqrt{2/\phi}$, $f = 1/(2 \pi \, \Delta x^{2} g^{3})$. 
 
The inner shock is positioned at $x=x_{{\rm is}}$. Once the integration has 
proceeded to this point the standard Rankine-Hugoniot jump conditions across 
a discontinuity $\Sigma$ must be satisfied:

\begin{equation}
[\rho u]_{\Sigma} = 0; \hspace{4mm} [\rho v^{2} + p]_{\Sigma} = 0; 
\hspace{4mm} [\frac{1}{2} v^{2} + \frac{\gamma}{\gamma - 1} 
\frac{p}{\rho}]_{\Sigma} = 0
\end{equation}

\noindent The pre-shock flow has an effectively
infinite Mach number and the strong shock jump conditions apply. 
If the post-shock values for the density and the velocity are given by 
$f_{2}$ and $g_{2}$, the Rankine-Hugoniot conditions for a 
$\gamma = 4/3$ gas specify that the post-shock 
value of the thermal energy density, $h_{2}$, is:

\begin{equation}
h_{2} = \frac{9}{2} f_{2} (g_{2} - g_{{\rm s}})^{2}
\end{equation}

\noindent where $g_{\rm s} = 7x_{\rm is}/(10 + \lambda)$ is the 
velocity of the inner shock relative to the star.

Once the post-shock conditions have been calculated, integration
proceeds until the contact discontinuity (CD) is reached. This occurs once the
flow velocity is equal to the coordinate velocity (\ie $v={\rm d}R/{\rm d}t$),
at which point $g(x=x_{\rm cd}) = 7x/(10+\lambda)$ is satisfied. At the CD,
we must also satisfy conservation of momentum. This requires that the 
pressure inside the bubble is equal to the impulse on the swept up shell:

\begin{equation}
\label{eq:cons_mtm}
4 \pi r^{2} p = \frac{{\rm d}(M_{{\rm sh}}v)}{{\rm d}t} = 
v\frac{{\rm d}M_{{\rm sh}}}{{\rm d}t} + 
M_{\rm sh}\frac{{\rm d}v}{{\rm d}t}
\end{equation} 

\noindent where here $M_{\rm sh}$ is the mass of the swept-up shell 
($M_{\rm sh}=4\pi\rho_{0}r_{\rm cd}^{3+\beta}/(3+\beta)$), and 
${\rm d}M_{\rm sh}/{\rm d}t = 4 \pi r^{2} \rho v$. 
Eq.~\ref{eq:cons_mtm} reduces to

\begin{equation}
\label{eq:cons_mtm2}
\rho_{0} r^{\beta} v^{2} + \frac{\rho_{0} r^{1+\beta}}{3+\beta} 
\frac{{\rm d}v}{{\rm d}t} = p
\end{equation} 

\noindent We now construct a dimensionless constant, $\theta$, 
from $Q$, $\dot{E}$, and $\rho_{0}$:

\begin{equation}
\label{eq:theta_form}
\theta \equiv \frac{Q^{-1/(10+\lambda)} \dot{E}^{1/(10+\lambda)}}
{\rho_{0}^{-1/(5+\beta)} \dot{E}^{1/(5+\beta)}}
\end{equation} 

\noindent It provides a measure of the ratio of the evaporated mass to the
swept-up mass. Inserting the similarity variables into Eq.~\ref{eq:cons_mtm2}
and substituting Eq.~\ref{eq:theta_form}, we find after some rearrangement:

\begin{equation}
\theta = \left(\frac{(\gamma-1) h}{x^{\beta}\left[g^{2} - \frac{3+\lambda}
{(10+\lambda)(3+\beta)} xg\right]}\right)^{1/(5+\beta)}_{\rm cd} 
\end{equation} 

\noindent The value of $\theta$ is evaluated once the integration stops
at the CD.

\subsection{Scale Transformation and Normalization}
\label{sec:scale_trans}
The dimensionless Eqs.~\ref{eq:ode1_evap}-\ref{eq:ode3_evap}
(and their counterparts for the 
USW region) are invariant under the following
transformation, which we shall call a normalization 
(\cf Chi\`{e}ze \& Lazareff \cite{CL1981}):

% NOTES: To produce a large bracket you must put a `\' after the \left or
% \right commands. To make an invisible bracket simply put a `.'
\begin{equation}
\label{eq:scale_trans}
\left.
\begin{array}{ll}
x \rightarrow & \alpha x \\
f \rightarrow & \alpha^{5+\lambda} f \\
g \rightarrow & \alpha g \\
h \rightarrow & \alpha^{7+\lambda} h 
\end{array}
\right\}
\end{equation}

\noindent These can be combined to obtain the normalizations for the
mass, and for the kinetic and thermal energies, of the bubble: 
%Several integrals related respectively to the mass, kinetic and 
%internal energy of the bubble are:

\begin{equation}
m = 4 \pi \int f x^{2} {\rm d}x 
 \hspace{15mm} [\alpha^{8+\lambda}] 
\end{equation}

\begin{equation}
k = 4 \pi \int \frac{1}{2} f g^{2} x^{2} {\rm d}x 
\hspace{10mm} [\alpha^{10+\lambda}] 
\end{equation}

\begin{equation}
i = 4 \pi \int h x^{2} {\rm d}x 
\hspace{17mm} [\alpha^{10+\lambda}]  
\end{equation}
%\end{eqnarray}

\noindent The power of $\alpha$ in the square brackets indicates
how each integral scales in a normalization. The full equations for the 
mass, kinetic and thermal energies, in the bubble are

\begin{equation}
\label{eq:mbub}
M_{\rm b} = 4 \pi \, \alpha^{8+\lambda} \, \Xi \, t \int^{x_{\rm cd}}_{x=0} f 
x^{2} \, {\rm d}x = \alpha^{8+\lambda} \frac{\dot{M}_{\rm c} t}{\phi} m
\end{equation}

\begin{equation}
\label{eq:ke_bub}
KE_{\rm b} = 2 \pi \, \alpha^{10+\lambda} \, \dot{E} t 
\int^{x_{\mathrm cd}}_{x=0} f g^{2}
x^{2} \, {\rm d}x = \alpha^{10+\lambda} \, \dot{E} t \, k
\end{equation}

\begin{equation}
\label{eq:ie_bub}
IE_{\rm b} = 4 \pi \, \alpha^{10+\lambda} \, \dot{E} t \int^{x_{cd}}_{x=0} h
x^{2} \, {\rm d}x = \alpha^{10+\lambda} \, \dot{E} t \, i
\end{equation}

\noindent The mass in the USW (SSW) region, $M_{\rm usw}$ ($M_{\rm ssw}$),
is obtained through the replacement of $m$ in Eq.~\ref{eq:mbub} with 
$m_{\rm usw}$ ($m_{\rm ssw}$), where the integral 
is evaluated from $x=0$ to $x=x_{\rm is}$ (from $x=x_{\rm is}$ to 
$x=x_{\rm cd}$). The total mass carried from the star by its wind is

\begin{equation}
M_{\rm w} = \int \dot{M}(t) \, {\rm d}t= \frac{10+\lambda}{16+3\lambda} 
\dot{M}_{\rm c} \, t
\end{equation}

\noindent The degree of mass-loading in the bubble, $\Phi_{\rm b}$, can be 
measured by the ratio $M_{\rm b}/M_{\rm w}$. Starting with the expression 
for $M_{\rm sh}$, (\cf text following Eq.~\ref{eq:cons_mtm}), using 
Eq.~\ref{eq:sim_r} to express $r_{\rm cd}$ in terms of $x_{\rm cd}$, 
employing Eq.~\ref{eq:theta_form} to
rewrite $\rho_{0}$ in terms of $\theta$, and using Eq.~\ref{eq:mdotc} and
the definition of $\Xi$, we find

\begin{equation}
\label{eq:m_sh}
M_{\rm sh} = \frac{4 \pi}{3+\beta} \, \alpha^{8+\lambda} \, \theta^{5+\beta} 
\, x_{\rm cd}^{3+\beta} \, \frac{\dot{M}_{\rm c} t}{\phi}
\end{equation}

\noindent The ratio of the swept-up mass in the shell 
to the bubble mass is hence:

\begin{equation}
M_{\rm sh}/M_{\rm b} = \frac{4 \pi \, \theta^{5+\beta} \, x_{\rm cd}^{3+\beta}}
{(3+\beta) \, m}
\end{equation}

\noindent The kinetic energy of the shell is $KE_{\rm sh} = \frac{1}{2}
M_{\rm sh}v_{\rm cd}^{2}$. Using Eqs.~\ref{eq:m_sh},~\ref{eq:edot},
\ref{eq:scale_trans}, and~\ref{eq:ode1_evap} through~\ref{eq:ode3_evap} we
obtain

\begin{equation}
\label{eq:ke_sh}
KE_{\rm sh} = \frac{2 \pi}{3+\beta} \alpha^{10+\lambda} \dot{E} t \, 
\theta^{5+\beta} \, x_{\rm cd}^{3+\beta} g_{\rm cd}^{2}
\end{equation}

\noindent We can also calculate the $pdV$ work done on the shocked interclump
medium in compressing it into a negligibly thin shell. The initial
energy per unit mass of the interclump gas immediately after passing through
the forward shock is:

\begin{equation}
\label{eq:energ}
E_{1} = \frac{1}{2}v_{1}^{2} + \frac{3}{2}\frac{p_{1}}{\rho_{1}}
\end{equation}

\noindent where here $v_{1}$ is the post-shock flow velocity 
($=\frac{3}{4} v_{\rm cd}$) and $p_{1}$ and $\rho_{1}$ are the post-shock
pressure and density. If we assume that the pressure and velocity of the
surrounding medium are both negligible, the post-shock pressure is
$p_{1} = \frac{3}{4}\rho_{0}r^{\beta}_{\rm cd} v_{\rm cd}^{2}$.
Substituting for $p_{1}$ into Eq.~\ref{eq:energ}, we find that the sum 
of the kinetic and thermal energies per unit mass behind a 
strong shock is $E_{1} = \frac{9}{16}v_{\rm cd}^{2}$. From our 
earlier assumptions, the swept up gas ends up with a velocity $v=v_{\rm cd}$ 
(\ie it is accelerated) and no internal energy. Hence, the final energy 
per unit mass is $E_{2} = \frac{1}{2}v_{\rm cd}^{2}$. Therefore the energy 
radiated away is $E_{rad} = E_{1}-E_{2} = \frac{1}{16} v_{\rm cd}^{2}$ per unit
mass. The total energy radiated away over the age of the bubble is:

\begin{equation}
\label{eq:e_rad}
E_{\rm rad} = \int 4\pi r_{\rm cd}^{2} (\rho_{0} r_{\rm cd}^{\beta} 
v_{\rm cd}) \frac{1}{16} v_{\rm cd}^{2} \, {\rm d}t
\end{equation}

\noindent where the integral is evaluated from $t=0$ to $t=t$.
Just as we obtained Eq.~\ref{eq:m_sh}, we find that

\begin{equation}
E_{\rm rad} = \frac{\pi}{4} \, \alpha^{10+\lambda} 
\, x_{\rm cd}^{2+\beta} \, g_{\rm cd}^{3} 
\, \theta^{5+\beta} \, \dot{E} t
\end{equation}

\noindent Global energy conservation demands that $\dot{E} t = 
KE_{\rm b} + IE_{\rm b} + KE_{\rm sh} + E_{\rm rad}$. With the 4 terms on 
the RHS scaling as $\alpha^{10+\lambda}$, the correct normalization of the
solution may be determined.

Finally, we can also calculate the fraction of mass lost by the
star over its lifetime that has yet to pass through the inner shock:

\begin{equation}
\rho_{\rm sw} = \frac{M_{\rm usw}}{M_{\rm w}} = \alpha^{8+\lambda} 
\left(\frac{16+3\lambda}{10+\lambda}\right) \frac{m_{\rm usw}}{\phi}
\end{equation}

\subsection{Solution Procedure}
Three parameters fully determine the form
of a given similarity solution ($\lambda$, $\phi$ and $x_{\rm is}/x_{\rm cd}$),
whilst a further parameter, ($\theta$), obtained only after a particular 
solution has been found, is a measure of the ratio of mass evaporated from the
clumps to the mass swept-up from the interclump medium. For bubbles whose 
evolution is significantly altered by mass-loading, we require that 
simultaneously both $\phi$ and $\theta$ are low.

The similarity equations were integrated with a fifth-order accurate
adaptive step-size Bulirsch-Stoer method using polynomial 
extrapolation to infinitesimal step size. Once the CD was reached, rescaling
was implemented with the relationships in Eq.~\ref{eq:scale_trans} so
that $x_{\rm cd} =1$. The mass, and kinetic and thermal energies
of the bubble were calculated, as were the kinetic energy of the shell
and the energy radiated from it. The correct normalization to satisfy
global energy conservation was then obtained. Finally, for given values of
$\dot{E}$, $Q$, and $t$, $x, f, g$ and $h$ may be
used to calculate the physical variables $r$, $\rho$, $u$, and $\varepsilon$.

\section{Results}
\label{sec:results}

\begin{figure*}
\begin{center}
\psfig{figure=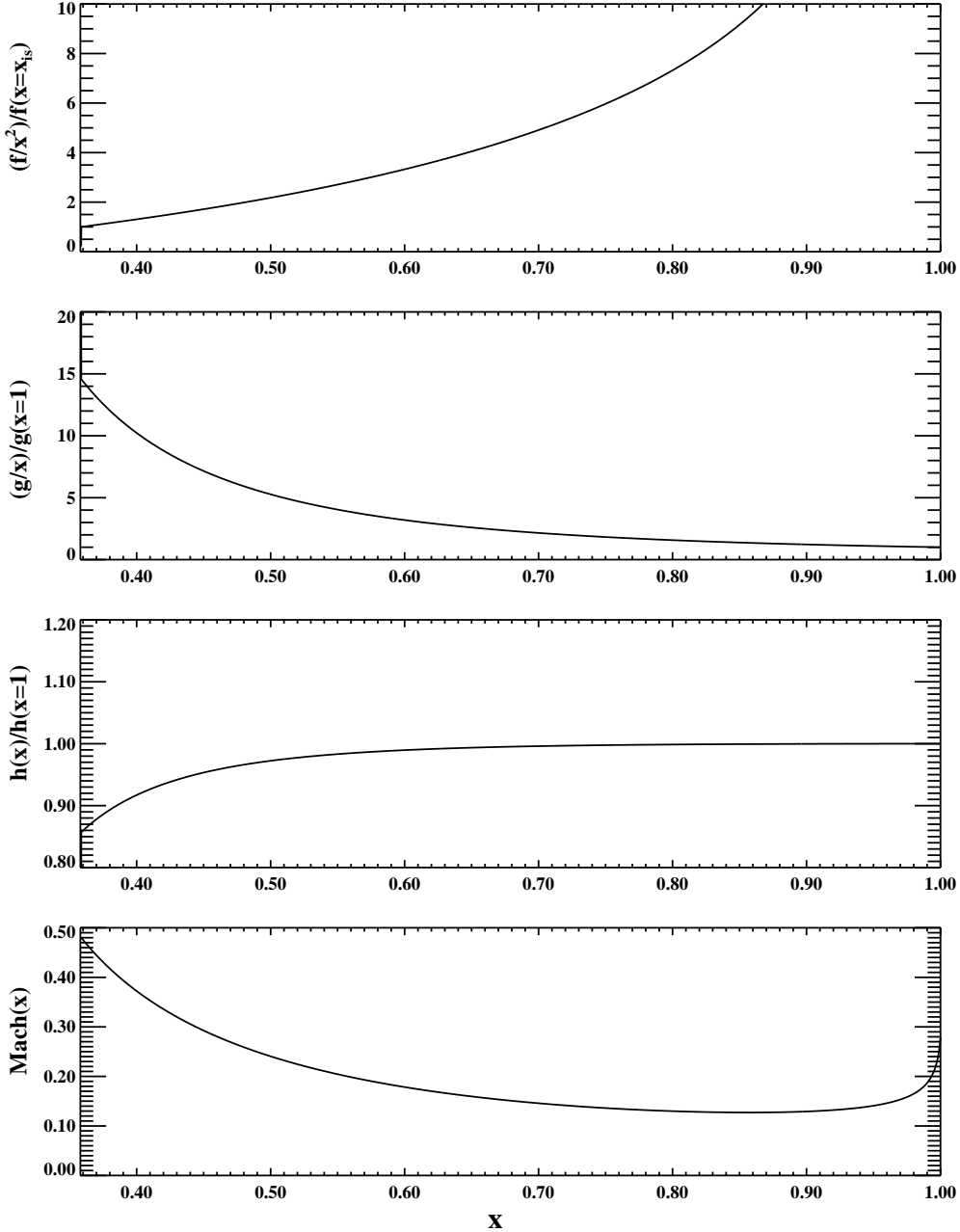,width=13.0cm}
\end{center}
\caption[]{Results for $\lambda = -3$, $x_{\rm is}/x_{\rm cd} = 0.359$, and 
$\phi = 50,000$. In the limit of large $\phi$ (\ie negligible mass-loading),
the earlier results of Dyson (\cite{D1973}) are recovered. Specific
combinations of the variables are chosen in order to facilitate comparison
with results given in Fig.~2 of Dyson (1973). In our work, $f$ is a measure
of density, $g$ is a measure of velocity, and $h$ is a measure of the thermal
energy density. Mach($x$) is the Mach number of the flow as a function of $x$.}
\label{fig:johns_results}
\end{figure*}

We first consider similarity solutions 
obtained by Dyson \& de Vries (\cite{DDV1972}) and 
Dyson (\cite{D1973}). These earlier works concerned the dynamical 
effects of a high velocity stellar wind incident on a smooth ambient
medium and contain no results for mass-loaded flows.
Dyson \& de Vries originally computed the form of the shocked
ambient gas behind the forward shock by parameterizing the cooling function,
which demanded that the ambient density, $\rho \propto r^{q}$ with $q=-1$.
They discovered that the shocked ambient gas cooled rapidly, and that
this region was very thin compared with the overall radius of the bubble.
Therefore, Dyson (\cite{D1973}) discarded the shocked ambient region 
in favour of a strong isothermal shock coincident with the contact 
discontinuity, which relieved the restriction on $q$. A central 
assumption for both papers was that the velocity of the stellar wind
was constant, which led to the stellar mass-loss rate varying as 
$\dot{M} \propto t^{q+2}$. Thus for an ambient density 
$\rho \propto r^{-2}$, the
energy input by the stellar wind was constant as a function of time. Since
in our work the stellar wind velocity is constant if the ambient density
falls off as $r^{-2}$ (\ie $\beta = -2, \lambda = -3$), and a constant
rate of energy input is a central assumption, we can make
a direct comparison between Dyson's (\cite{D1973}) results with those 
obtained here.

\begin{figure}
\begin{center}
\psfig{figure=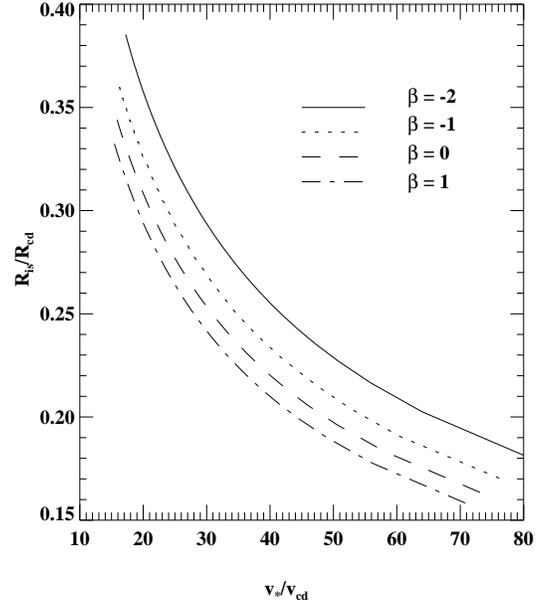,width=9.5cm}
\end{center}
\caption[]{The ratio of the radii of the inner and outer shocks as a 
function of $g_{*}/g_{\rm cd}$ for cases with negligible mass loading. 
For $\beta=-2$ the earlier results of Dyson (\cite{D1973}) are again
recovered (\cf his Fig.~3). For $\beta\neq 2$ the velocity 
of the stellar wind is a function of time, so these curves cannot be 
compared to those in Dyson's Fig.~3. For $\beta = -2,-1,0,1$, the
corresponding values of $\lambda$ are $-3,-\frac{2}{3},\frac{5}{3},4$.}
\label{fig:shk_pos}
\end{figure}

In Fig.~\ref{fig:johns_results} we show results for $\lambda = -3$ with
negligible mass-loading (large $\phi$). We chose specific combinations 
of the variables in order to facilitate comparison with results given in
Fig.~2 of Dyson (\cite{D1973}): agreement is excellent.
In Fig.~\ref{fig:shk_pos} we plot the relative
radii (or velocities) of the inner and outer shocks as a function 
of $g_{*}/g_{\rm cd}$, the
ratio of the stellar wind velocity to the velocity of the contact
discontinuity (or outer shock). These calculations were again made with
negligible mass-loading. The results for the $\beta=-2$ case may again
be directly compared to the earlier results of Dyson (his Fig.~5).
In Fig.~\ref{fig:shk_pos} we also show results for other values of $\beta$.
These cannot be compared to Dyson's earlier results because when $\beta \neq 2$
the stellar wind velocity is no longer constant with time. We find that 
as the value of $\beta$ increases (which also corresponds to the 
mass-loading becoming increasingly dominant at large radii), the 
shocked region broadens for a given ratio of $g_{*}/g_{\rm cd}$.

\begin{figure*}
\begin{center}
\psfig{figure=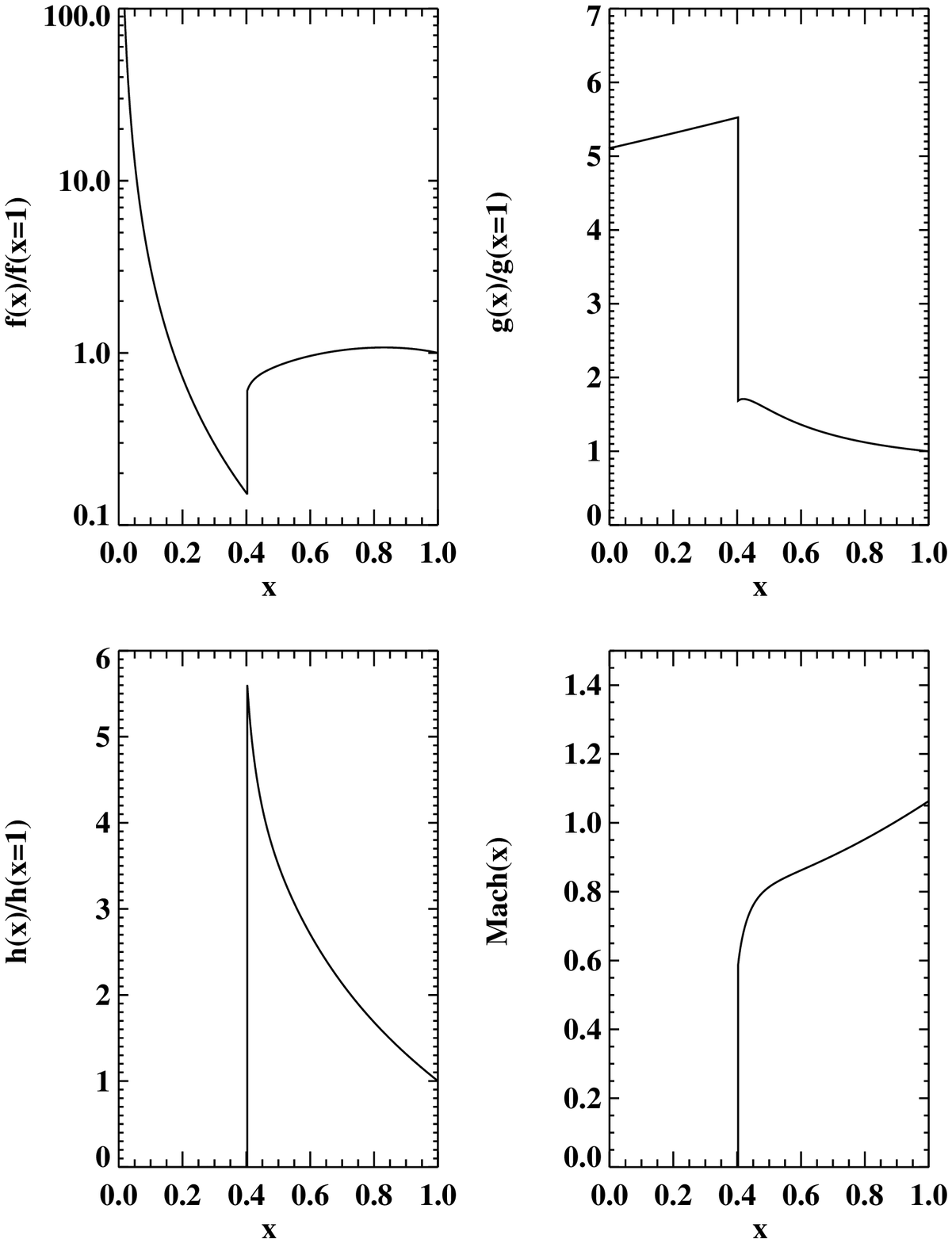,width=12.0cm}
\end{center}
\caption[]{Results for a bubble with significant mass-loading. This 
particular solution is for $\lambda=4$, $\phi=0.199$, and 
$x_{\rm is} = 0.403 x_{\rm cd}$. The bubble mass is 8.4 times the
total mass lost by the star. The ratio of the swept-up mass to the
bubble mass is $M_{\rm sh}/M_{\rm b} = 0.54$.}
\label{fig:large_ml}
\end{figure*}

In Fig.~\ref{fig:large_ml} we present results for
a case in which mass-loading dominates the evolution of the bubble. In
this particular case, $\lambda = 4$, $\phi = 0.2$, 
$x_{\rm is} = 0.403 x_{\rm cd}$, and $\theta = 0.62$. The bubble mass is
over 8 times higher than the total mass lost by the star, and almost
twice the mass of the ambient medium swept up into the thin
shell. 30 per cent of the mass lost by the star has still to pass
through the inner shock. The high degree of mass-loading increases the
post-shock density, and dramatically reduces the temperature in the
post-shock region as more and more mass is 
evaporated. This reduces the sound speed, and there is eventually
a sonic transition close to the CD as the flow becomes supersonic in 
the frame of the clumps. 

Values of several parameters calculated from solutions with $\lambda=-3$ 
are given in Table~\ref{tab:lambda-3}. If the mass-loading is negligible
(illustrated in the table by $\phi=50,000$), we find the following: i) as the 
ratio of $x_{\rm is}/x_{\rm cd}$ increases, the fraction of energy contained
in the form of thermal energy in the bubble and that in the form of 
kinetic energy of the shell decreases, whilst that in the form of kinetic 
energy of the bubble increases. The fraction of energy radiated away
by the shocked interclump gas stays relatively constant, and is always a 
small part ($<6$ per cent) of the total mechanical energy of the 
stellar wind over its lifetime. ii) The mass contained in the bubble is larger
than the mass of the swept-up shell only when the relative radius of the
inner shock is large. 
iii) For small $x_{\rm is}/x_{\rm cd}$, about 90 per cent of the total
mechanical energy of the stellar wind is split approximately evenly between
the thermal energy contained in the SSW region and the the kinetic energy
of the swept up shell. As $x_{\rm is}/x_{\rm cd}$ increases, these fractions
decrease at roughly the same rate, whilst the kinetic energy of the
gas inside the bubble increases, until it eventually may exceed 90 per 
cent of the total energy. iv) The kinetic energy of the swept-up shell
is always as great or greater than the thermal energy of the bubble (this
is also true for the solutions in Table~\ref{tab:lambda-3} with $\lambda=-3$,
$\phi=1.0$).

\begin{table*}
\begin{center}
\caption{Properties of solutions with $\lambda=-3$ ($\beta=-2$). 
We consider two cases: negligible mass-loading ($\phi=50,000$) and
moderate mass-loading ($\phi=1.0$). Solutions with $\lambda=4$ and
$\phi=0.2$ are also listed. From left to right, the columns
indicate: i) the ratio of the shock radii, ii-v) the fraction of energy 
in the form of thermal energy and kinetic energy in the bubble,
kinetic energy in the swept-up shell, and radiated energy, 
vi) the value of $\theta$, 
vii) the ratio of mass within the bubble to the total mass lost by the star,
viii) the ratio of mass in
the swept-up shell to mass in the bubble, ix) the fraction of the
mass lost by the star which has yet to pass through the inner shock.}
\label{tab:lambda-3}
\begin{tabular}{lllllllll}
\hline
$x_{\rm is}/x_{\rm cd}$ & $IE_{\rm b}$ & $KE_{\rm b}$ & $KE_{\rm sh}$ & 
$E_{\rm rad}$ & $\theta$ & $\Phi_{\rm b}$ & $M_{\rm sh}/M_{\rm b}$ & 
$\rho_{\rm sw}$ \\
\hline
$\lambda=-3, \phi = 50,000$ & & & & & & & & \\
0.34 & 0.45 & 0.03 & 0.47 & 0.06 & 1286 & 1.075 & 162 & 0.018 \\
0.49 & 0.40 & 0.09 & 0.46 & 0.06 & 603 & 1.068 & 35.6 & 0.053 \\
0.69 & 0.26 & 0.29 & 0.40 & 0.05 & 247 & 1.052 & 5.8 & 0.175 \\
0.89 & 0.05 & 0.74 & 0.19 & 0.02 & 86 & 1.046 & 0.60 & 0.505 \\ 
\hline
$\lambda=-3, \phi = 1.0$ & & & & & & & & \\
0.12 & 0.47 & 0.002 & 0.47 & 0.06 & 45 & 2.64 & 4098 & 0.0008 \\
0.26 & 0.46 & 0.01 & 0.47 & 0.06 & 9.7 & 2.26 & 219 & 0.0079 \\
0.53 & 0.36 & 0.15 & 0.44 & 0.05 & 1.97 & 1.63 & 12.3 & 0.079 \\
0.73 & 0.21 & 0.40 & 0.35 & 0.04 & 0.85 & 1.20 & 2.86 & 0.232 \\
0.95 & 0.01 & 0.92 & 0.05 & 0.01 & 0.20 & 1.00 & 0.11 & 0.724 \\
\hline 
$\lambda=4, \phi= 0.2$ & & & & & & & & \\
0.18 & 0.70 & 0.005 & 0.24 & 0.06 & 15.3 & 1.26 & 4151 & 0.0079 \\
0.37 & 0.66 & 0.05 & 0.23 & 0.06 & 3.24 & 1.36 & 171 & 0.077 \\
0.42 & 0.49 & 0.38 & 0.11 & 0.03 & 0.66 & 7.87 & 0.71 & 0.303 \\
\hline
\end{tabular}
\end{center}
\end{table*}

\begin{figure*}
\begin{center}
\psfig{figure=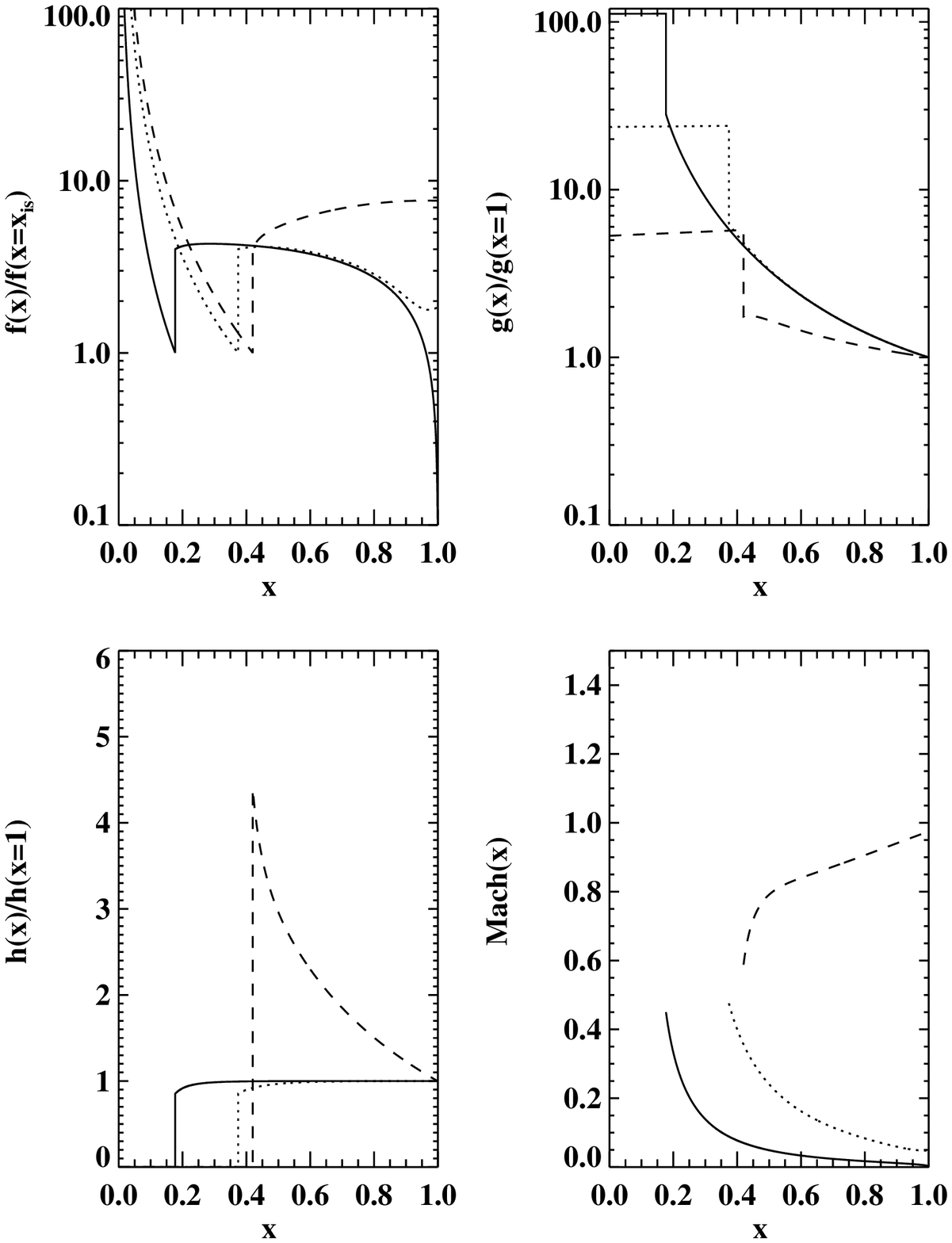,width=12.0cm}
\end{center}
\caption[]{Results for bubbles in which the radial dependence of the 
mass-loading is $\propto r^{4}$.
Three solutions with different values of $x_{\rm is}/x_{\rm cd}$ are shown:
0.177, 0.374, and 0.420 (solid, dotted, and dashed lines). 
$\Phi_{\rm b}$ is equal to 1.26, 1.36 and 7.86
respectively, whilst $M_{\rm sh}/M_{\rm b}$ is equal to 4151, 171, 0.71
respectively. For ease of comparison, $f$
is scaled such that its pre-shock value is unity in each case.}
\label{fig:lambda4_var}
\end{figure*}

When we set $\phi=1.0$, the bubble is just beginning to become mass-loaded
for small values of $x_{\rm is}/x_{\rm cd}$. However, for $\lambda=-3$, the 
swept-up mass dominates the bubble mass in this regime, as can be
seen from the values in Table~\ref{tab:lambda-3}. We find that
solutions for which {\em both} $\Phi_{\rm b}$ is large 
{\em and} $M_{\rm sh}/M_{\rm b}$ is small exist only if $\lambda \gtsimm 4$.  
Solutions for $\lambda=4$ ($\beta=1$) and $\phi=0.2$ are illustrated in 
Fig.~\ref{fig:lambda4_var}. The solution with $x_{\rm is}/x_{\rm cd} = 0.420$
is strongly mass-loaded, and in contrast to the solutions with $\lambda=-3$ the
mass of the swept up shell is also less than $M_{\rm b}$. We note that when
$x_{\rm is}/x_{\rm cd}$ is varied, notable differences between the profiles 
of $f$, $g$, and $h$ occur in Fig.~\ref{fig:lambda4_var}. 
For instance, when the bubble is only weakly mass-loaded
the density at the CD is smaller than immediately after the inner shock. 
However, once the bubble is more 
strongly mass-loaded the density continuously rises from the inner shock
to the CD. Other differences include less deceleration of the flow,
a sharp fall in the temperature (instead of a gradual rise), and a rise in
the Mach number. Finally, when the bubble becomes strongly
mass-loaded, the fractional distribution of the total energy between
$IE_{\rm b}$, $KE_{\rm b}$, $KE_{\rm sh}$, and $E_{\rm rad}$ for a given 
$x_{\rm is}/x_{\rm cd}$ ratio is substantially different than when the 
mass-loading is negligible. This is hardly surprising given the change
seen in the profiles of the flow variables.

\begin{figure*}
\begin{center}
\psfig{figure=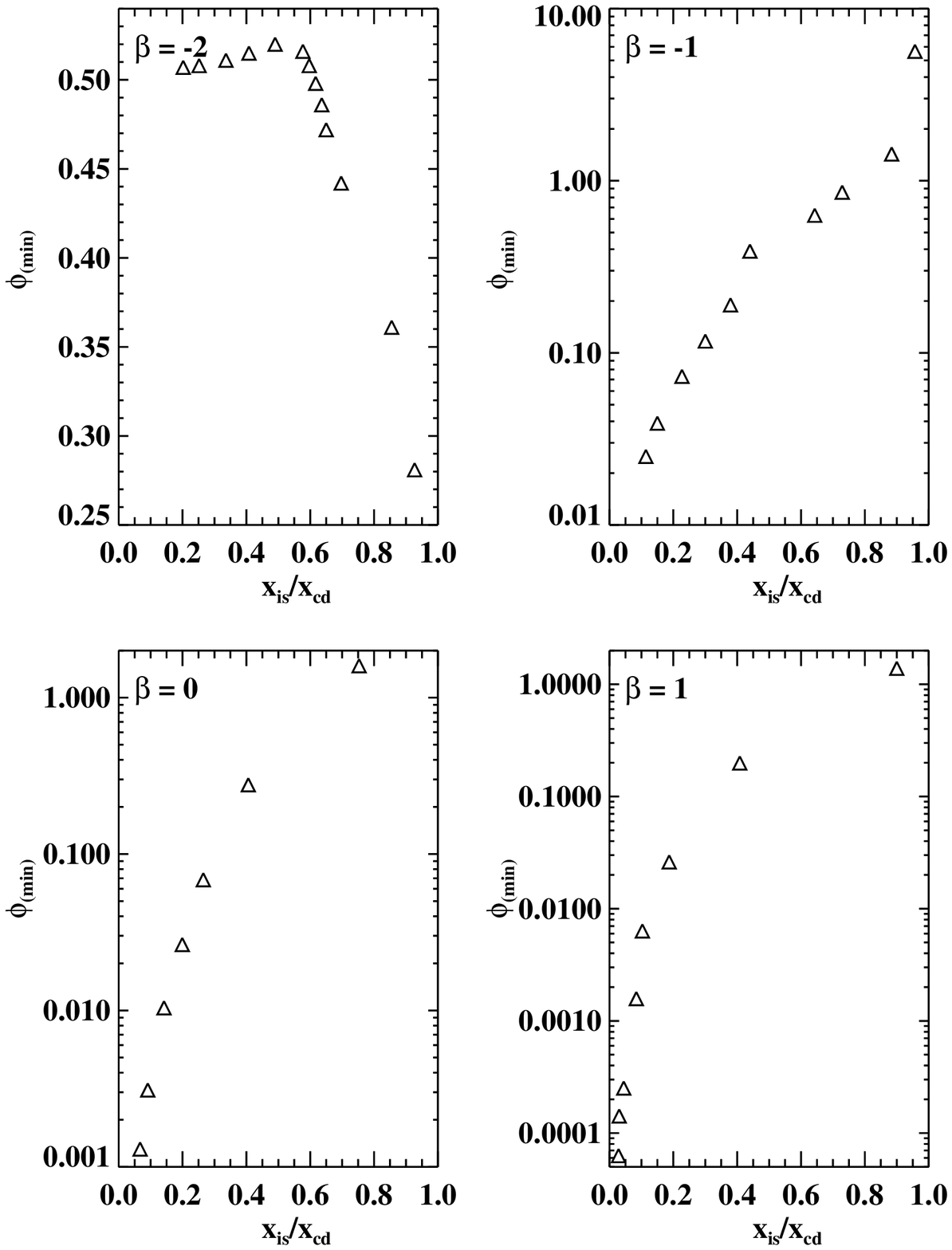,width=12.0cm}
\end{center}
\caption[]{The minimum value of $\phi$ (\ie corresponding to 
the maximum possible mass-loading) for which solutions could be found, 
as a function of $x_{\rm is}/x_{\rm cd}$ and
$\beta$ (or, alternatively $\lambda$).}
\label{fig:phi_min}
\end{figure*}

An important finding is that for a given $\lambda$
and ratio $x_{\rm is}/x_{\rm cd}$, there exists a minimum value of $\phi$ at
which we can find a solution. In other words, {\em there is a limit 
to the maximum mass-loading which the bubble can undergo}. 
In Fig.~\ref{fig:phi_min} we show the value of $\phi_{min}$ for various
values of $x_{\rm is}/x_{\rm cd}$ and $\beta$. This finding has the 
corollary that for a given value of $\phi$ there is a maximum value of 
$x_{\rm is}/x_{\rm cd}$.
The numerical reason that we are unable to obtain solutions with 
$\phi < \phi_{min}$ is that the denominator of the derivatives from 
Eqs.~\ref{eq:ode1_evap}-\ref{eq:ode3_evap} approaches zero {\em without} the
numerators doing likewise. The physical reason is that there is a 
straightforward feedback mechanism limiting the amount of mass-loading that 
may take place: as mass from the clumps is evaporated, the temperature of the
shocked region falls, which reduces the rate at which further mass can
be evaporated. Similarly, if the mass-loading were suddenly inhibited, 
the temperature of the shocked region would rise, leading to an increase 
in the evaporation rate. There is no reason to believe that actual wind
bubbles would be at this limit because the degree of mass-loading simply
depends on the number of clumps present and their individual mass 
injection rates (which in our model is characterized by the value of 
$\phi_{b}$). For a given mass, the number density of clumps, 
$n_{c} \propto r_{c}^{-3}$ where $r_{c}$ is the clump radius, whilst
the mass injection rate from each clump $\dot{m}_{c} \propto r_{c}$.
Thus we can only comment that for small clumps, it is more likely that
this maximum mass-loading limit will actually occur.

\begin{figure*}
\begin{center}
\psfig{figure=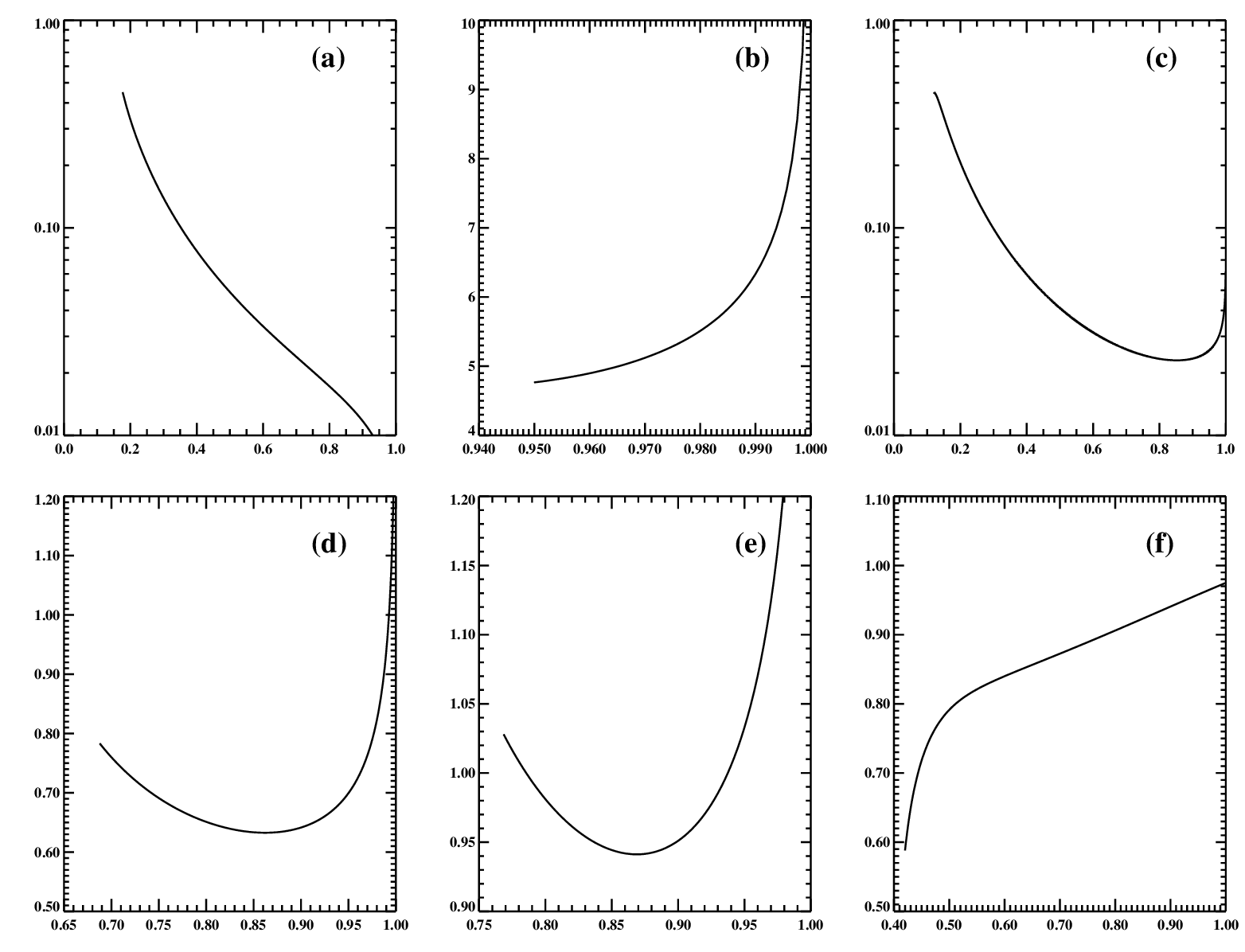,width=17.0cm}
\end{center}
\caption[]{The Mach number as a function of $x$ for a number of 
different solutions. With respect to the clumps, the shocked region 
can be either entirely subsonic, entirely supersonic, or have up to
two sonic points. The corresponding solution shown in each panel is:
{\bf a} $\lambda = 4$, $\phi=0.2$, $x_{\rm is}/x_{\rm cd} = 0.177$;
{\bf b} $\lambda = -3$, $\phi=1.0$, $x_{\rm is}/x_{\rm cd} = 0.950$;
{\bf c} $\lambda = -3$, $\phi=1.0$, $x_{\rm is}/x_{\rm cd} = 0.120$;
{\bf d} $\lambda = -3$, $\phi=50,000$, $x_{\rm is}/x_{\rm cd} = 0.688$;
{\bf e} $\lambda = -3$, $\phi=50,000$, $x_{\rm is}/x_{\rm cd} = 1.301$;
{\bf f} $\lambda = 4$, $\phi=0.2$, $x_{\rm is}/x_{\rm cd} = 0.420$.}
\label{fig:mach}
\end{figure*}

We further note that the radial profile of the Mach number of the 
flow relative to the clumps can take many different forms, depending 
on the input parameters used. In Fig.~\ref{fig:mach} we show several 
examples. The profile of the Mach number is influenced by the fact 
that mass-loading tends to drive the Mach number
towards unity, whilst the spherical divergence tends to drive it away from
unity (\cf Hartquist \etal \cite{HDPS1986}). Panel a) is for a case in 
which mass-loading is small. In this case the divergence term wins,
and drives the Mach number towards zero. Panel b) is for a similar
situation, but in this case because the inner shock is very close to the CD,
the initial postshock Mach number is greater than unity so the divergence
term drives it towards infinity. In panel c) we see that the divergence
is initially dominant, but as the flow approaches the CD the Mach number
is driven back towards unity. However, this behaviour is not due to the flow
mass-loading (although $\Phi_{\rm b} = 2.64$), but is rather because the
density increases near the CD whilst the pressure remains constant,
producing a decrease in the sound speed (the same effect occurs for the
flow shown in Fig.~\ref{fig:johns_results} which has no mass-loading). 
The same situation occurs in panel d) where the fall in sound speed
is great enough to drive the flow through a sonic 
transition. The flow corresponding to panel e) 
is very similar to that associated with panel d) except we now have 
two sonic transitions due to the initial post-shock flow being 
supersonic with respect to the clumps. In plot f) the
flow is significantly mass-loaded ($\Phi_{\rm b} = 7.86$), and the 
mass-loading term dominates the divergence term such that the Mach 
number is driven towards unity.

\begin{figure*}
\begin{center}
\psfig{figure=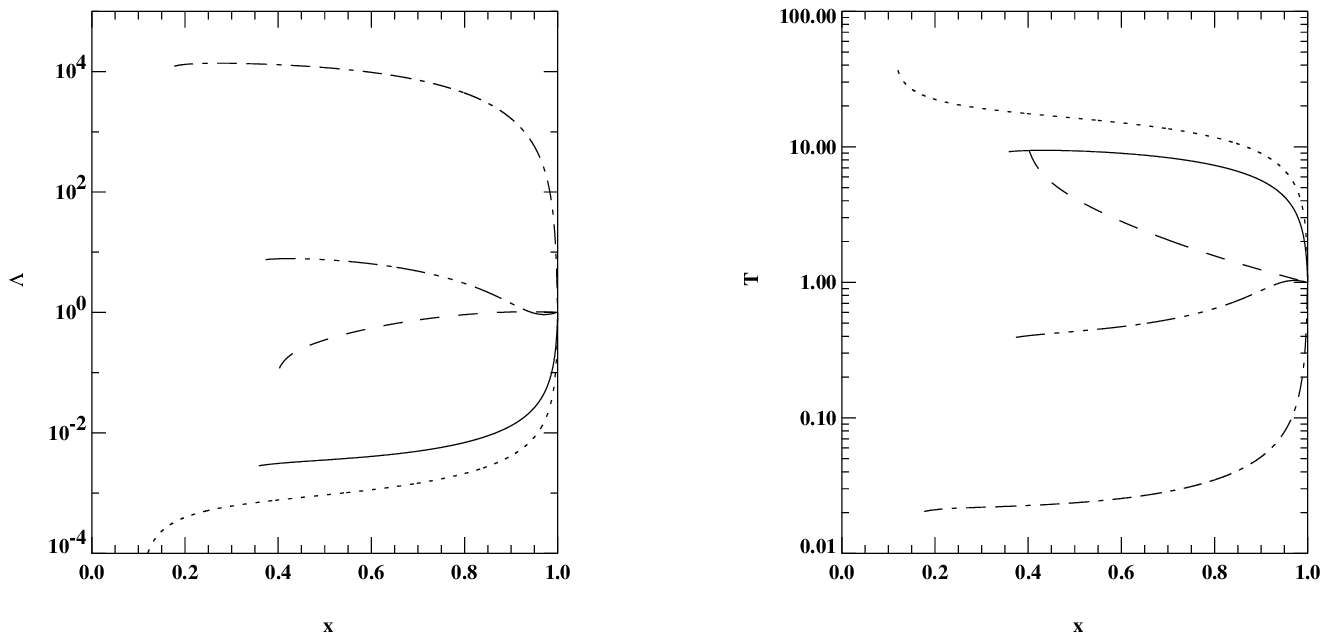,width=18.0cm}
\end{center}
\caption[]{Radial profiles of emissivity per unit volume and temperature, 
normalized to values of 1.0 at the limb, for a number of the simulations 
previously discussed. These are $\lambda =-3$, $\Phi_{\rm b}=1.0$ (solid); 
$\lambda=-3$, $\Phi_{\rm b}=2.64$ (dots); $\lambda=4$,
$\Phi_{\rm b}=1.26$ (dashes); $\lambda=4$, $\Phi_{\rm b}=1.36$ (dot-dash); 
$\lambda=4$, $\Phi_{\rm b}=7.87$ (dot-dot-dot-dash). The general trend is for
a reduction in the central emissivity and an increase in the central 
temperature with increased mass loading.}
\label{fig:xray}
\end{figure*}

Finally, we have calculated profiles of the X-ray emissivity as a function
of radius (see Fig.~\ref{fig:xray}). We assume that the emissivity 
$\Lambda \propto n^{2} T^{-1/2}$, which is a good approximation over the
temperature range $5 \times 10^{5} {\rm K} \ltsimm T \ltsimm 5 
\times 10^{7}{\rm K}$ (\cf Kahn \cite{K1976}). Also plotted in 
Fig.~\ref{fig:xray} are
the radial temperature profiles. A bubble un-affected by mass-loading 
expanding into a surrounding medium with an $r^{-2}$ profile of density has
a higher central temperature but a lower central emissivity than at its limb.
Conversely, if it is expanding into a medium with an $r^{1}$ profile of
density, the situation is reversed, as the central temperature is lower and
the central emissivity is higher than at its limb. If the bubble is 
mass-loaded, the general trend is for a reduction in the central emissivity
and an increase in the central temperature relative to the limb. To date, the
only stellar wind bubbles which have been successfully observed in X-rays
are two Wolf-Rayet ring nebulae, NGC~6888 (\eg Wrigge \etal \cite{WCMK1998})
and S308 (Wrigge \cite{W1999}). Both have X-ray luminosities lower than
expected from the standard model (Weaver \etal \cite{WMCSM1977}). This 
discrepancy is normally attributed to two possibilities : either conductive 
evaporation of gas from the cold dense outer shell into the bubble interior,
or enhanced cooling resulting from high metallicities in the cooling gas
(see MacLow \cite{M2000}). It is not surprising, therefore, that our
mass-loading simulations indicate that evaporation from clumps {\em within}
the bubble may also be compatible with current observations. The latest
X-ray satellites, {\it Chandra} and {\it XMM}, should provide significantly
improved observations which may enable us to distinguish between these 
competing mechanisms. One might also expect that they may allow
discrimination between a mass-loaded bubble expanding into an ambient medium
with $\rho \propto r^{1}$ and a bubble unaffected by mass-loading expanding
into a medium with $\rho \propto r^{-2}$.

\section{Summary}
\label{sec:conclusions}
We have investigated the evolution of a mass-loaded wind blown bubble 
with a constant rate of ejection of mechanical energy from a central 
stellar wind. Mass-loading is assumed to occur by the conductively induced 
evaporation of clumps in the region of shocked stellar wind. A central
assumption is that the shocked interclump medium
rapidly cools to a dense, negligibly thin, shell through which the
clumps can pass undisturbed. The requirement that the solution be
self-similar imposes a link 
between the radial variation of the interclump density  
($\rho \propto r^{\beta}$) and the mass-loading from the clumps 
($\dot{\rho} \propto r^{\lambda}$) which forces $\lambda = (5 + 7\beta)/3$.

We first reproduced solutions with negligible mass-loading and $\beta=-2$,
which we compared with results obtained by Dyson (\cite{D1973}). 
Excellent agreement for the structure of the bubble was found. We also 
found that the ratio of the  radii (velocities) of the inner and outer 
shocks decreases with increasing $\beta$ for a given ratio of the 
{\em current} stellar wind velocity to the CD velocity. 

We then investigated the changes in the structure of the bubble for 
different values of $\lambda$, $\phi$, and $x_{\rm is}/x_{\rm cd}$.
The central conclusions are:
\begin{itemize}
\item Substantial mass-loading of the wind-blown bubble can occur over
a wide range of $\lambda$. However, to additionally satisfy
the requirement that the bubble mass is larger than the mass of
the swept-up shell, large values of $\lambda$ are needed ($\lambda \gtsimm 4$).
\item The profiles of the flow variables are significantly altered under
conditions of large mass-loading. In particular, the average density 
of the shocked region is larger, the deceleration of the flow is shallower, 
and the temperature of the shocked wind rapidly decreases. 
\item The Mach number of the shocked stellar wind relative to the 
clumps which are injecting the mass can take several different forms.
The shocked region can be either entirely subsonic, entirely supersonic,
or have one or sometimes two sonic points.
\item For a given $\lambda$ and ratio of the shock radii, there exists
a maximum mass-loading that can occur.
\item Mass-loading tends to reduce the emissivity in the interior of the
bubble relative to its limb, whilst simultaneously increasing the central
temperature relative to the limb temperature.
\end{itemize}

\begin{acknowledgements}
JMP would like to thank PPARC for the funding of a PDRA position, and 
Sam Falle and Rob Coker for helpful discussions. We would also like
to thank an anonymous referee whose suggestions improved this paper.
\end{acknowledgements}

\end{document}